\documentclass[preprint]{elsarticle}
\usepackage[numbers]{natbib}
\usepackage{amsmath,amsfonts} 
\usepackage{amssymb}
\usepackage{bm}
\usepackage{bbm}

\usepackage{graphicx}

\newcommand{\be}{\begin{equation}}
\newcommand{\ee}{\end{equation}}
\newcommand{\ba}{\begin{eqnarray}}
\newcommand{\ea}{\end{eqnarray}}
\newcommand{\ban}{\begin{eqnarray*}}
\newcommand{\ean}{\end{eqnarray*}}

\newcommand{\vct}[1]{{\bf #1}}

\renewcommand\Im{\operatorname{Im}}

\begin{document}

\title{Fractal Interpretation of Bacteria Light Harvesters}

\author[df]{Julian Juhi-Lian Ting}
\ead{juhilian@gmail.com}
\ead[url]{http://amazon.com/author/julianting}
\address[df]{De-Font Research Institute, Taichung 40344, Taiwan, R.O.C.}

\begin{abstract}

Bacteria light harvesters are interpreted as Vicsek fractal,
based upon their morphology,
even though such fractals contain only one, or perhaps two, generations.
After using fractal dimensions to describe the geometry, we progress to
use connectivity matrices to make an improved description of the
light harvesters and make connections with our previous studies.


\end{abstract}



\maketitle

\section{introduction}

The study of protein as fractals has provided fruitful results.
In particular, Onsager pointed out the long-range interaction at
phase transitions is best described with fractals,
which resulted in resolving the Levinthal paradox of protein folding~\citep{Moret2001}.
These studies are, however, mostly on linear biopolymers, 
whereas there are many other examples that are not linear.
An immediate generalization of a linear chain to a loopless fractal
object is a Vicsek fractal~\citep{Blumen2004}.
Bacterial light harvesters, cancellous bone and tooth enamel are of such shapes.
The mechanical properties of bio-structures such as cancellous bone and
tooth enamel have been considered without mentioning their fractal properties~\citep{Wegst2015}.
In this paper we consider the optical and thermal properties inspired by
bacterial light harvesters using Vicsek fractals.

The structures of bacterial light harvesters (LH) have been known since 1995 at atomic resolution~\citep{Kuhlbrandt1995}.
Both the inner antenna (LH1) and the outer antenna (LH2) have a
toroidal shape, which are composed of the same modules.
The exact numbers of modules involved are variable~\citep{Kuhlbrandt1995,Karrasch1995,Bradforth1995}.
The outer antenna, LH2, is smaller and consists of nine units for
{\it Rhodopseudomonas acidophila}~\citep{McDermott1995,Papiz2003}
with an outer diameter $68$ \AA.
The inner antenna, LH1, is larger, as it contains the reaction center (RC),
and has 17 units, for {\it Blastochloris viridis},
which has an elliptic shape, and has an outer longer diameter $124.5$ \AA ~and a shorter axis $120.2$ \AA ~\citep{Qian2018}.
We list some of them in Table \ref{more}.
Each module is further composed of several carotenoids, $\alpha$-helical polypeptides and bacteriochlorophylls of various types,
but the number involved is smaller than the number of units for the ring.

The number of units surrounding a RC is similar for photosystem I and photosystem II of plants~\citep{Croce2011,Caffarri2014}.
The exact dimensions of each molecule can be read with software such as
Jmol or PyMOL from a .cif or .pdb file in the protein data bank (PDB,
http://www.rcsb.org/pdb/).

\begin{table}[tb]
\centering
\begin{tabular}{p{0.3\columnwidth}ccll}
\hline 
protein & PDB ID & symmetry & top view & side view\\
\hline \\


LH1-RC from {\it Thermochromatium tepidum}  ~\citep{Niwa2014}              & 4V8K & C16 &
\includegraphics[width=0.15\textwidth, angle=0]{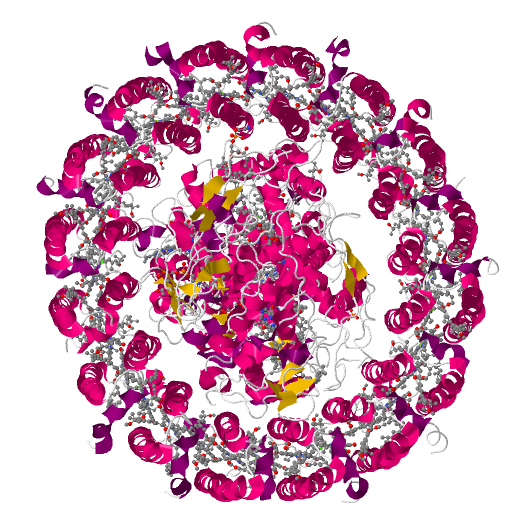} &
\includegraphics[width=0.15\textwidth, angle=0]{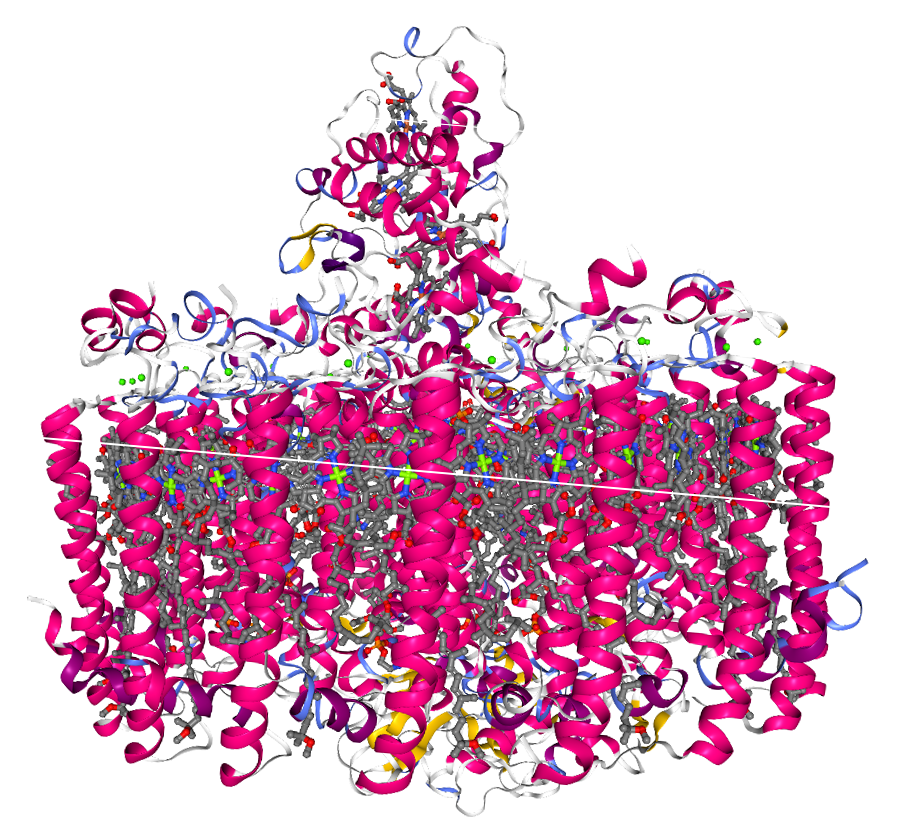} 
\\

LH1-RC from {\it Blastochloris viridis} ~\citep{Qian2018}              & 6ET5 & C17 &
\includegraphics[width=0.15\textwidth, angle=0]{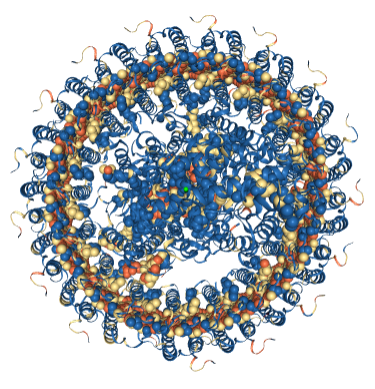} &
\includegraphics[width=0.15\textwidth, angle=0]{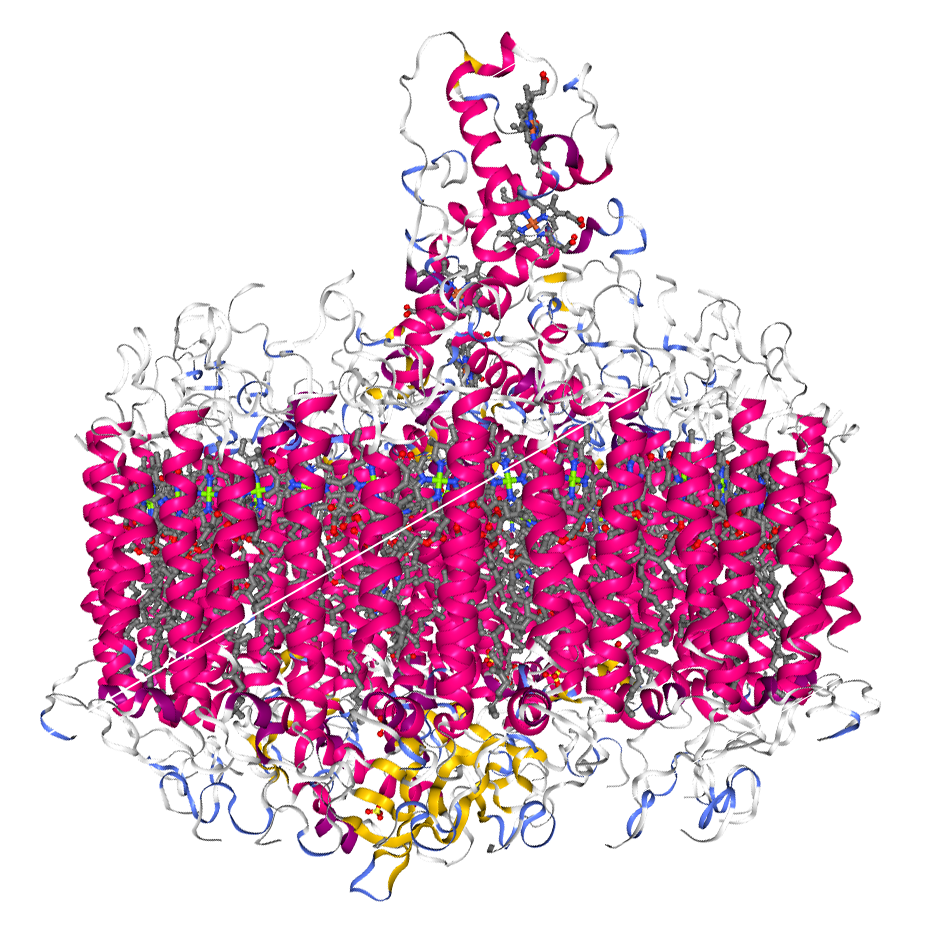} 
 \\


LH2 B800-850 from {\it Rhodospirillum molischianum} ~\citep{Koepke1996}              & 1LGH & C8 &
\includegraphics[width=0.15\textwidth, angle=0]{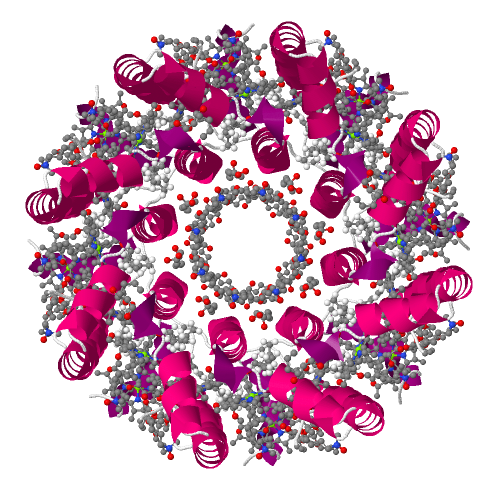} &
\includegraphics[width=0.15\textwidth, angle=0]{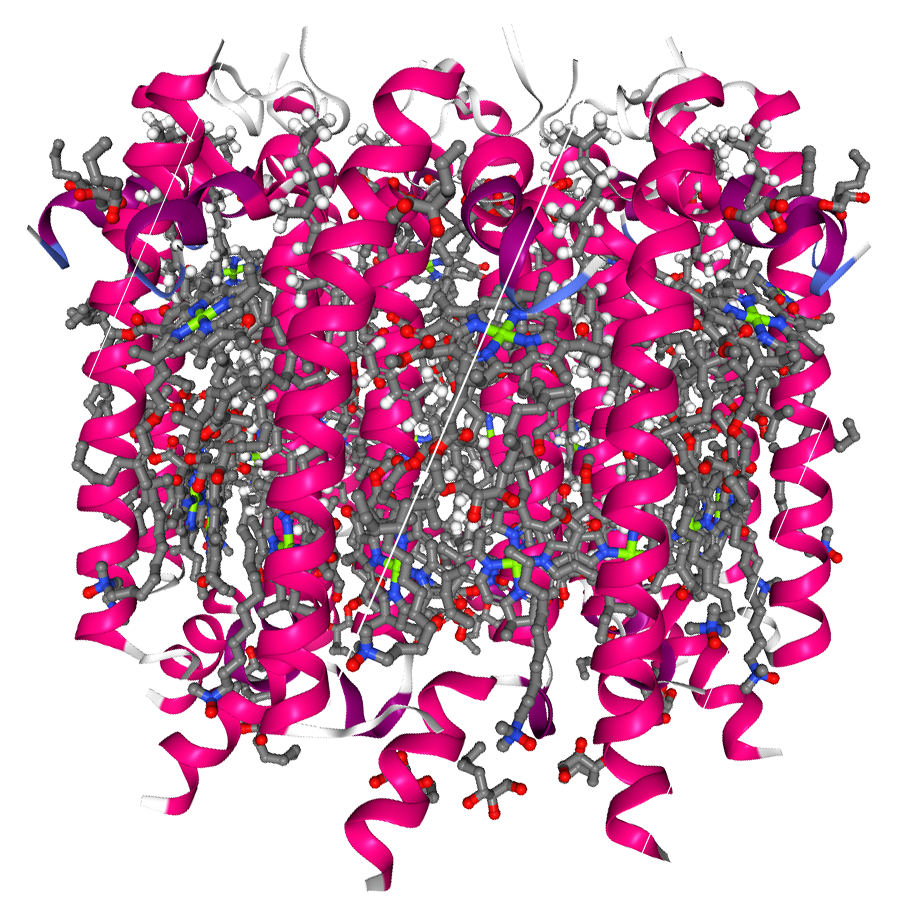} 
 \\

LH2 B800-850 from {\it Rhodopseudomonas acidophila} \citep{Papiz2003}              & 1NKZ & C9 &
\includegraphics[width=0.15\textwidth, angle=0]{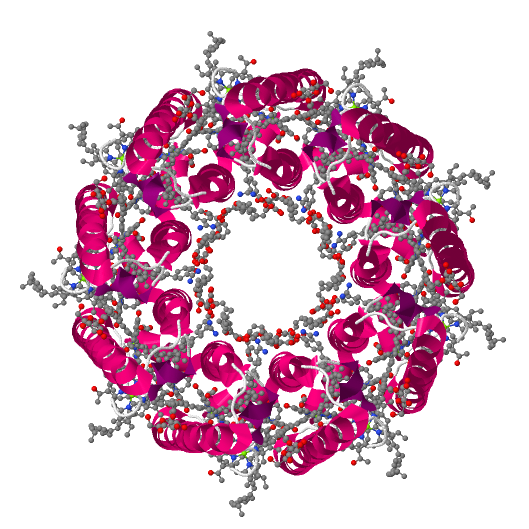} &
\includegraphics[width=0.15\textwidth, angle=0]{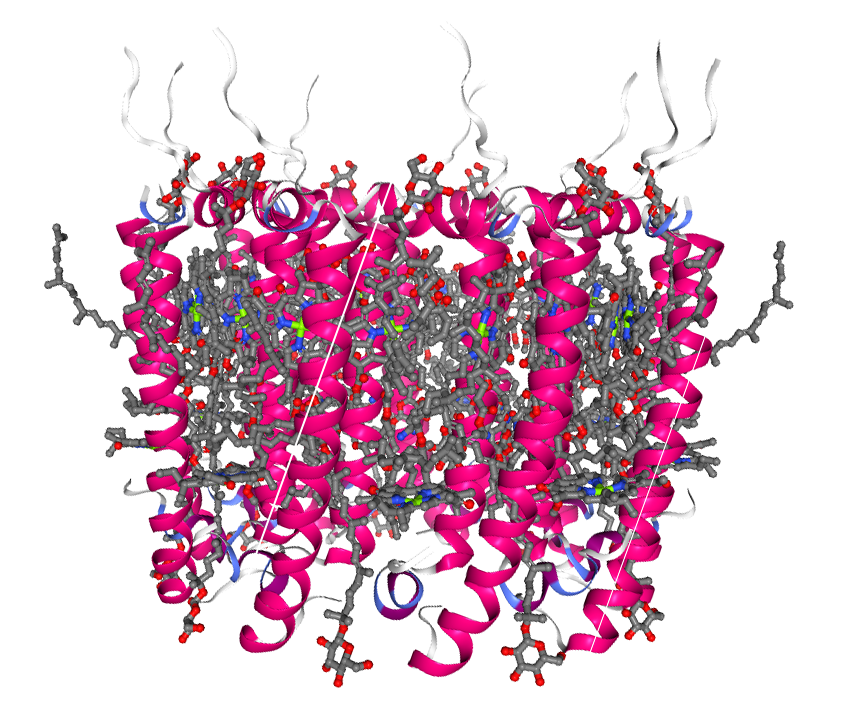}
\\

\hline 
\end{tabular}
\caption{Various bacteria light harvesters.}
\label{more}
\end{table}

The most salient feature of these molecules is their symmetry,
which is particularly important in physics.
The second important feature that attract physicists attention is
that they can interact with electromagnetic waves,
which make them either energy harvesters or heat radiators.


The tradition to call the photosynthetic light harvesting antennae, in English, is traceable to 1960s~\citep{Clayton1965},
even though nobody ever considered them as antennae seriously until we
modeled them as a simple loop antenna and provided sixteen physical
interpretations without {\it ad hoc} parameter~\citep{Ting2018,Ting2019,Ting2019d}.
These molecules are not made of conductor,
but they are neither dielectric,
because the dielectricity is defined for only bulk material~\citep{Jackson2007}.
Nature achieved the required electromagnetic properties through the geometry~\citep{Pendry1999},
as natural materials typically have few components that have poor intrinsic
properties but complicated architectures.
It is hence important to provide a method to describe the geometry of
the light harvesters and to obtain the desired property through that geometry.

Theories to describe the radiation properties of nanostructures have emerged since about year 2000~\citep{Novotny2012,Schuller2009}.
A nanoantenna is similar to metamaterial~\citep{Pendry2000,Pendry2004,Jahani2016,Monticone2017},
but the latter requires periodic structures, even though these two names are typically mixed~\citep{Paniagua-Dominguez2019}.
To be qualified as a nanoantenna the particle must have a size range from about $25~nm$ to $2~nm$~\citep{Stockman2011}.
Furthermore, for a radiator to be efficient the particle size must
be smaller than Wien's length (thermal wavelength),
$\hbar c/k_B T \approx 7.5~ \mu m$ at room temperature~\citep{Thompson2018}.


In the following, we provide two methods to describe the geometry or 
morphology of the light harvesters and 
obtain the power of radiation through scattering theory.
The first is a quasi-fractal description of the nanoantennae in terms of 
their fractal dimension; 
the second is a complete description of the nanoantennae using the connectivity 
matrix that provides a framework to calculate the performance of the 
radiator and results in a new interpretation of the band structure of the radiation.
The former is useful in experiments, whereas the latter is useful for theorists.

\section{Quasi-fractal interpretation of light harvesters}

Photosynthetic light harvesters have been considered as dendritic materials 
without further fractal interpretation~\citep{DAmbruoso2005},
but the Vicsek fractal resembles the partitioned ring shape of the bacterial 
light harvesters~\citep{Blumen2004}.
For instance, LH2 of 1LGH is a Vicsek fractal with a functionality $F=8$ 
at the first generation ($g=1$), 
which has a fractal dimension~\citep{Blumen2004}
\be
d_{1LGH} = \dfrac{\ln (F+1)}{\ln 3} =  \dfrac{\ln 3^2}{\ln 3} =2\,.
\ee
This number indicates that the light harvester of 1LGH has reached the 
most compact form in the two-dimensional world.
$d_{1NKZ}$ is slightly larger than two. 
The reasons partly arise from slight overlap of the sub-units, 
partly because a cylinder is only quasi-two-dimensional, and still
partly because we are using a formula for an infinite generation of 
a fractal whereas a real light harvester has only one generation.
But the deviation cannot be too large; no larger ring is found.

The LH1-RC of 6ET5 has functionality $F=17$ at $g=1$, and
\be
d_{6ET5} = \dfrac{\ln (F+1)}{\ln 3} =  \dfrac{\ln (2 \times 3^2)}{\ln 3} =2 + \dfrac{\ln 2}{\ln 3} = 2 + \ln_3 2\,.
\ee
The fact that $2<d_{6ET5} <3$ indicates that LH1-RC is almost a three-dimensional object.
The reason is that such a light harvester is not composed of the same unit.
In particular, the unit at the centre, i.e. the reaction centre, differs from the unit composed of the ring as can be
seen clearly from the side view of table \ref{more}.
The functionality for 4V8K is $16$, which also results in $2<d_{4V8K} <3$.

LH2 made of the same unit is surely a two-dimensional cylinder, whereas
LH1 has its centre made of another unit and is
surely of different length at the centre to become a three-dimensional object.
We insert only the functionality into the formula; the results obtained,
i.e., $d_{1LGH}$, $d_{1NKZ}$, $d_{6ET5}$, and $d_{4V8K}$,
are consistent with the image obtained from experiments, which indicate the
validity of using a Vicsek fractal to model the light harvesters.

Each unit considered here is further composed of several carotenoids,
$\alpha$-helical polypeptides, and bacteriochlorophylls;
they typically form a complex of C3 symmetry.
The number of such sub-units is less than the number of units involved forming the light harvesters, 
which make their fractal dimension slightly larger than unity.
This effect is one further reason for our interpreting them as fractals;
it is the second generation of the fractal considered.

Because our fractal contains only one, or perhaps two as pointed out in the previous paragraph, generation,
the number of particles involved and the radius of gyration are trivial.
The fractal dimension is, however, a measure of the spatial arrangement of nanoparticles,
which is valid even though the fractal contains only one generation.
We hence connect the radiative properties with the fractal dimension.
A consequence of being fractal is that they exhibit enhanced optical responses~\citep{Stockman1996}.
The critical indices of the enhancement factor are determined by the optical spectral dimension
of the fractal~\citep{Stockman1992}.
Antennae of such geometry have been analyzed~\citep{Shan2004}.
Fractality seems to be the secret of rapid signal processing in the primary step of photosynthesis.

A real Vicsek fractal has zero volume but an infinite surface area as
its generation approaches infinity,
which means that nature is using less material to build the light harvester
even though nature cannot approach the mathematical limit.
Such structural design is applicable for materials that require light weight and stiffness, but simultaneously strength and
toughness, and economy in engineering.
Similar two-dimensional structures exist in cancellous bone and tooth enamel even though these materials appear three-dimensional~\citep{Wegst2015}.

As many viral proteins can be self-assembled under suitable conditions
to form  macromolecules,
these light harvesters might also exhibit power-law distributions near
criticality~\citep{Phillips2014},
but the number of light harvesters in PDB and the range of their size
distribution are insufficient for such an analysis.

\section{Connectivity Matrix Description}

The most important issue of our concern is the power of energy transfer,
whether harvesters are receivers or radiators.
There are at least two methods to obtain the power in the literature:
the first method assumes initially a dipole moment to obtain
the Clausius-Mossotti relation that connects
polarizability and the relative permittivity of a
material~\citep{Nikbakht2017}.
This approach is inadequate because, as mentioned in the
introduction, the available information is the geometry rather than the polarizability of the material.
The second method uses scattering theory to obtain the power radiated,
hence requires only the geometry of the material.
With the reflection and transmission coefficients,
the refractive index and impedance can be obtained.
Hence effective permittivity and permeability can be defined~\citep{Smith2002}.
Both methods require a Green's tensor to describe the geometry.

The  Green's tensor has other names in the literature, 
such as connectivity matrix, transfer-rate matrix or interaction matrix.
It is a complex-symmetric matrix, i.e.
${\hat{\mathbb G}}_{ij}={\hat{\mathbb G}}_{ji}$ but ${\hat{\mathbb G}} \ne {\hat{\mathbb G}}^\dagger$, in its most general form.
Its eigenvectors are hence not orthogonal in general.

If the dissipation effects are not taken into account, 
the complex elements become real.
The matrix proposed in our previous paper can be normalized to fit such purposes~\citep{Ting1999b}.
The matrix elements have normally value unity if the elements $i$ and $j$ are connected, and zero, otherwise.
Such a matrix is generally studied in discrete mathematics, in particular graph theory, and is called an adjacency matrix.

Many dynamical properties of the antenna, 
such as the spectra and the relaxation modes,
can be obtained from the spectrum of the eigenvalues or eigenvectors of this matrix.
In particular, the spectral radius of eigenvalues of $\hat{\mathbb G}$ determines the width of the frequency band
of which the transmission probability is maximum.
The width of the spectrum decreases on increasing the distance $d$ between two elements,
because of weak coupling between particles at large distances.
If these eigenvalues were obtained numerically, they would seem to have no other distinction,
but if the characteristic equation were factorized algebraically, we should find modes
within the same factor grouped to form bands of the radiation spectrum~\citep{Ting1999b}.
The number of factors is hence equal to the number of bands involved.

Suppose the object has a homogeneous temperature $T_{obj}$ placed in
vacuum whereas the environment is at temperature $T_{env}$. 
In equilibrium, $T_{obj}=T_{env}=T$, the autocorrelation function $C$
of the electric field is related to the imaginary part of the dyadic
Green's function $G_{ij}$ of the object by the
fluctuation-dissipation theorem~\citep{Rytov1978, Eckhardt1984},
\begin{equation}\label{kruger1}
\begin{split}
C_{ij}^{eq}(T)&
\equiv\left\langle E_i(\omega;\textbf{r}) E_j^*(\omega;\textbf{r}')\right\rangle^{eq}
\equiv 
{\left\langle\vct{E}(\omega;\textbf{r}) \otimes \vct{E}^*(\omega;\textbf{r}')\right\rangle^{eq}_{ij}}\\
&= \dfrac{c^2}{\omega^2} \left[a_T(\omega)+a_0(\omega)\right] \textrm{Im}
G_{ij}(\omega;\textbf{r},\textbf{r}'),\
\end{split}
\end{equation}
where $\otimes$ denotes a dyadic product, and
$a_T(\omega)\equiv (4\pi)^2 \omega^4 \hbar  /(c^4\exp[\hbar \omega/k_BT]-1)$ describes the thermal contribution to quantum fluctuations.
The zero point fluctuations do not contribute to heat radiation. 
We will denote $\mathbb{G}\equiv G_{ij}(\omega ;\textbf{r},\textbf{r}')$ in the following.

Because the electric fields obey the Helmholtz equation
\begin{equation}
\left[ \mathbb{H}_0+\mathbb{V}-\frac{\omega^2}{c^2}\mathbb{I}\right]\vct{E}=0 \;,\label{eq:H2}
\end{equation}
the Green's function is the solution of
\begin{equation}
\left[ \mathbb{H}_0+\mathbb{V}-\frac{\omega^2}{c^2}\mathbb{I}\right]\mathbb{G}=\mathbb{I}.\label{eq:H}
\end{equation}
In the above two equations, the first term $\mathbb{H}_0=\boldsymbol{\nabla}\times\boldsymbol{\nabla}\times$ describes the free space, 
whereas $\mathbb{V}=\frac{\omega^2}{c^2}(\mathbb{I}-\epsilon+\boldsymbol{\nabla}\times\left(\frac{1}{\mu}-\mathbb{I}\right)\boldsymbol{\nabla}\times)$ 
is the potential introduced by the object.
We are assuming isotropic and local material, therefore $\epsilon$ and $\mu$ are scalars.
$\mathbb{G}_{0}$ is the Green's function of free space. 

With
$\textrm{Im}
\mathbb{G}=-\mathbb{G}\textrm{Im} \mathbb{G}^{-1}\mathbb{G}^*$ and
$\textrm{Im} \mathbb{V}=
\textrm{Im}
(\mathbb{G}^{-1}-\mathbb{G}_{0}^{-1})$ from Eq.~\eqref{eq:H}~\citep{Eckhardt1984}, 
we obtain
\begin{align}\label{kruger2}
C^{eq}(T)&=C_0+C(T)-a_T(\omega) \frac{c^2}{\omega^2}\mathbb{G}\textrm{Im} \mathbb{G}_0^{-1} \mathbb{G}^*,\\
C(T_{obj})&=-a_{T_{obj}}(\omega)\frac{c^2}{\omega^2}\mathbb{G}\notag\textrm{Im} \mathbb{V}\mathbb{G}^*\\&=-a_{T_{obj}}\frac{c^2}{\omega^2}(\omega) \int\limits_{\rm obj}d^3r' d^3r'' G_{ij}(\vct{r},\vct{r}')\notag\\&\times\textrm{Im} V_{jk}(\vct{r}',\vct{r}'') G_{kl}^*(\vct{r}'',\vct{r}'''),\label{eq:CT}
\end{align}
where $C_0= \frac{c^2}{\omega^2} a_0(\omega)\textrm{Im} \mathbb{G}$ is
the zero point term.  
Equation~\eqref{kruger2} is the equilibrium of two terms at finite temperature:
Firstly, $C(T)$ contains an explicit integral over the sources within the radiator.
As $\Im \mathbb{V}$ is only nonzero inside the radiator, which we shall calculate according to specific radiator.
Secondly, the third term in Eq.~\eqref{kruger2},
is the contribution from the environment because
\begin{equation}
C^{env}(T_{env})=-\frac{c^2}{\omega^2} a_{T_{env}}(\omega)\mathbb{G} \textrm{Im}
\mathbb{G}_0^{-1} \mathbb{G}^* \,.\label{eq:Cenv}
\end{equation}

Our heat sink, the outer space, can be considered as a very large black cavity maintained at temperature $T_{env}$.
With the identity
\begin{equation}
\mathbb{G}=\mathbb{G}_0-\mathbb{G}_0\mathbb{T}\mathbb{G}_0\label{eq:GT}.
\end{equation}
we obtain
\begin{align}\label{coldobject}
&C^{env}(T_{env})=\langle \vct{E}_s\otimes \vct{E}_{s}^*\rangle=(1-\mathbb{G}_0\mathbb{T})\langle \vct{E}_0\otimes\vct{E}^*_0\rangle\notag\\&
\times(-\mathbb{T}^*\mathbb{G}^*_0+1)
=-a_{T_{env}}(\omega)\frac{c^2}{\omega^2}\mathbb{G}\textrm{Im}
\mathbb{G}_0^{-1} \mathbb{G}^* \;,\
\end{align}
in which 
$\mathbb{T}$ is the scattering amplitude of the radiator~\citep{Tsang2000, Rahi2009}.

If $T_{env} \approx 0$ we arrived at the heat radiation solution.
The heat radiation of the object at temperature $T_{obj}$ can be obtained from Eq.~\eqref{kruger2} for $C(T_{obj})$,
\begin{align}
&C(T_{obj})=a_{T_{obj}}(\omega) \frac{c^2}{\omega^2}\textrm{Im} \mathbb{G}-C^{env}(T_{obj})\label{eq:radfin},
\end{align}
where $\mathbb{G}$ is found using Eq.~\eqref{eq:GT}.

It is not necessary to derive all the terms in Eq.~\eqref{kruger2} because
the explicit expression for $C(T_{obj})$ in Eq.~\eqref{eq:CT} contains the Green's function with one argument inside and one argument outside the object. 
While this function can be in principle derived, 
it is more convenient to express $C(T_{obj})$ in terms of the Green's function with both arguments outside the object, 
as it is directly linked to the scattering operator by Eq.~\eqref{eq:GT}. 
$C^{env}$ has all the sources outside the object and hence can be found in terms of this Green's function, which 
is already obvious in Eq.~\eqref{coldobject}.
A more rigorous way to derive $C^{env}$ is also possible:
The environment sources, described by  $\varepsilon_{env}$, 
can be thought of as being everywhere in the infinite space complementary to the object, 
infinitesimal in strength (environment ``dust'' \citep{Eckhardt1984}), i.e. $\varepsilon_{env}\to1$.
$C^{env}$ in Eq.~\eqref{eq:Cenv} can hence be written
\begin{align}
&C^{env}(T_{env})\notag=a_{T_{env}} (\omega)\\&\lim_{\varepsilon_{env}\to1}\int_{\rm outside}d^3
r'\widetilde{G}_{ik}(\textbf{r},\textbf{r}')\textrm
{Im}\varepsilon_{env}
\widetilde{G}^*_{jk}(\textbf{r}'',\textbf{r}') \;.\label{cenv}
\end{align}
Here, we introduced a Green's function $\tilde{\mathbb{G}}$ with $\mathbb{V}$ inside the object and
$\varepsilon_{env}$ outside. This is a simple modification of
$\mathbb{G}$ as a finite $\varepsilon_{env}-1$ only changes the
external speed of light so that $c$ in $\mathbb{G}$ is replaced by
$c/\sqrt{\varepsilon_{env}}$.

If our only interest is the total power, 
the first term, i.e., the equilibrium field need not be derived, 
as it contains no Poynting vector.

\section{Summary}

In the present paper two methods are juxtaposed to describe the morphology of the
heat radiators, i.e., fractal dimension and connectivity matrix.
The former is a scalar whereas the latter is a dyadic.
The former allows a succinct description of the radiator whereas
the latter allows a detailed calculation of the performance of the radiator.
The former interpretation is original, hence the title of this paper,
whereas the latter is generally found in the literature,
but we indicate that some other methods are not useful and a connection
to our previous studies is given.




\bibliography{c:/dna/library} 

\end{document}